\documentclass[12pt]{iopart}
\usepackage{graphicx}

\begin{document}

\title{Generalised Longitudinal Susceptibility for Magnetic Monopoles in Spin Ice}

\author{Steven T. Bramwell}

\address{London Centre for Nanotechnology and Department of Physics and Astronomy, University College London, 17-19 Gordon Street, London WC1H 0AJ, UK.}

\date{today}

\begin{abstract}
The generalised longitudinal susceptibility $\chi({\bf q}, \omega)$ affords a sensitive measure of the spatial and temporal correlations of magnetic monopoles in spin ice. Starting with the monopole model, a mean field expression for  $\chi({\bf q}, \omega)$ is derived as well as 
expressions for the mean square longitudinal field and induction at a point.  Monopole motion is shown to be strongly correlated, and both spatial and temporal correlations are controlled by the dimensionless monopole density $x$ which defines the ratio of the magnetization relaxation rate and the monopole hop rate. Thermal effects and spin lattice relaxation are also considered. The derived equations are applicable in the temperature range where the Wien effect for magnetic monopoles is negligible. They are discussed in the context of existing theories of spin ice and the following experimental techniques: dc and ac-magnetization, neutron scattering, neutron spin echo, and longitudinal and transverse field $\mu$SR.  The monopole theory is found to unify diverse experimental results, but several discrepancies between theory and experiment are identified. One of these, concerning the neutron scattering line shape, is explained by means of a phenomenological modification to the theory. 
\end{abstract}

\maketitle

\newpage

\section{Introduction}

Following the paper of Castelnovo, Moessner and Sondhi~\cite{CMS} on emergent magnetic monopoles, there has been renewed interest in the properties of spin ice~\cite{Harris,BramwellHarris, Ramirez, BramwellGingras}. Magnetic monopole currents  were first envisaged by Ryzhkin~\cite{Ryzhkin} while Jaubert and Holdsworth~\cite{Jaubert, Jaubert2} studied the closely related problem of magnetic relaxation in spin ice by means of numerical simulations of the dipolar spin ice model~\cite{Siddharthan,Hertog} and of a dual monopole electrolyte. Evidence for the characteristic non-Ohmic conductivity signature of a weak monopole electrolyte - the Wien effect - was reported in Refs. \cite{Nature, NatPhys}, where the term `magnetricity' was coined. 

Experimental evidence indicates that magnetic monopoles afford an economical description of spin ice at temperatures below $\sim 10$ K~\cite{Fennell, Morris, Kadowaki}. In particular, down to about 0.3 K the equilibrium specific heat is well described by Debye-H\"uckel theory~\cite{Morris,Zhou,CMS2}. However, to account in detail for the monopole currents and magnetic relaxation is generally a tricky problem, especially in the regime of slow dynamics at subkelvin temperatures, and this is an ongoing subject of investigation~\cite{NatPhys, Klemke, Matsuhira-new}. 

Prior to the recent wave of interest, the spin dynamics of spin ice were studied in detail by Matsuhira {\it et al.}~\cite{Mats} and Snyder {\it et al.}~\cite{Snyder} by ac-susceptibility, by Ehlers {\it et al.}~\cite{Ehlers1,Ehlers2} using neutron spin echo, by Lago {\it et al.}~\cite{Lago} using muon spin relaxation ($\mu SR$), by Orend\'{a}\v{c} {\it et al.}~\cite{Kitagawa} using bulk magnetocalorimetric methods, by Kitagawa {\it et al.}~\cite{Kitagawa} using nuclear quadrupole resonance, and by Sutter {\it et al.}~\cite{Sutter} using nuclear forward scattering. Previous studies of the spatial spin correlations by neutron scattering may be found, for example, in Refs. ~\cite{Harris,Bramwell1,Kanoda,Fennell1, Fennell2,Yavorskii,Clancy}.  

The spin correlations in the spin ice state are characterised by two remarkable features~\cite{Fennell, Morris,Chang, Chang2}. The first is a property common to many ice-type models, that transverse magnetization (or polarisation) fluctuations are essentially unrestricted while longitudinal fluctuations are strongly suppressed at low temperature. This behaviour is captured in the phenomenological theory of Youngblood and Axe~\cite{YA} (formulated to describe ice rule ferroelectrics), in which the deconfined defects do not carry any Coulombic charge. The second remarkable property, of course, is that in spin ice these defects do carry a magnetic Coulomb charge~\cite{CMS,Ryzhkin}. However, they are also associated with a `Dirac string network' of spin configurations, that while not pairing the monopoles, does restrict their motion to some extent~\cite{Jaubert,CMS2,Quench}.  

Any complete theory of spin ice needs to account for the difference between longitudinal and transverse correlations, the Coulombic interactions of monopoles and the effect of the Dirac string network. However, different experiments may pick out one or another of these three properties, so approximate models are useful. If monopoles are the focus then it is of most interest to discuss the longitudinal response as this is a highly sensitive measure of the spatial and temporal correlation of magnetic monopoles, as shown below. The simplest approach to treating the monopole correlations is a `magnetolyte' model of freely diffusing magnetic charges, in which the effect of Dirac strings is subsumed into the transport coefficients, and electrolyte properties such as Debye-H\"uckel screening~\cite{CMS2}, Bjerrum pairing~\cite{Nature, Zhou,CMS2} and the Wien effect~\cite{Nature,NatPhys} may be naturally formulated. Another (and earlier) approach~\cite{Ryzhkin} accounts for the ignored spin degrees of freedom in the form of an effective reaction field, and this approach has recently been developed to include magnetic charge screening~\cite{Ryzhkin-new}.  

The aim of the present work is to calculate a generalised longitudinal susceptibility $\chi({\bf q}, \omega)$ for magnetic monopoles in spin ice and to explore its application to experiment. The theory described here is only a modest extension of the earlier approaches of Ryzhkin~\cite{Ryzhkin} and Castelnovo {\it et al.}~\cite{CMS2}, but one that is necessary for the purpose of comparing theory with experiment. A useful by-product of this work is a clarification of the relationship between these two approaches, and their relation to that of Youngblood and Axe~\cite{YA}. The equations discussed here are valid at temperatures that are sufficiently high to avoid the non equilibrium physics of the Wien effect~\footnote{In spin ice the direct current Wien effect takes the form of a transient increase in charge density under applied field, before equilibration to a value lower than the zero field value~\cite{Vojtech}; the alternating current Wien effect should take the form of a steady state increase in current.} for magnetic monopoles ($>0.4 $ K for Dy$_2$Ti$_2$O$_7$)~\cite{Nature}, but not so high that high energy relaxation processes become important $(> 10$ K for Dy$_2$Ti$_2$O$_7$)~\cite{CMS2}.  
The results of the present paper are applicable to zero and weak applied field ($\mu_0 H \ll 1$ T)~\footnote{A detailed discussion of the dynamical susceptibility of spin ice in the vicinity of the critical point induced by a field of $\mu_0 H \approx 1$ T applied along the $[111]$ axis ~\cite{Sakakibara} has been given by Shtyk and Feigel'man \cite{Shtyk}.}. 

\section{Ryzhkin's Approximation}

Ryzhkin~\cite{Ryzhkin} explored the monopole dynamics of spin ice by applying the Jaccard theory of water ice defects in the water ice-spin ice analogy.  He showed that the magnetic current density ${\bf J} = {\bf J}_++{\bf J}_-$ is related to the rate of change of magnetization ${\bf M}$ by the equation (in our notation): 
\begin{equation} \label{flux}
{\bf J} =  \frac{\partial {\bf M}}{\partial t}
\end{equation}
and he derived the rate of entropy production associated with the flow of the magnetic charges:
\begin{equation}\label{entropy}
\left(T \frac{\partial S}{\partial t}\right)_{\rm irreversible} =  \mu_0 {\bf J} \cdot ({\bf H} - \chi_T^{-1}{\bf M}),
\end{equation}
where $S$ is entropy and ${\bf H}$ is magnetic field. 
He finally used these relations to 
derive a thermodynamic equation of motion: 
\begin{equation}\label{Ryzhkin}
{\bf J} =   \kappa ({\bf H} - \chi_T^{-1}{\bf M}).
\end{equation}
Here $\kappa = u c Q \mu_0$ is the monopole conductivity, $u = u_+ = -u_-$ is the monopole mobility, $c = c_+ + c_-$ is the total concentration of free monopoles and $Q=Q_+ = -Q_-$ is the monopole charge. The isothermal susceptibility is predicted to be 
$\chi_T = 2\chi_C$ where $\chi_C$ is the nominal Curie susceptibility for the spin ice system $(\chi_C \approx 3.95/T$ for Dy$_2$Ti$_2$O$_7$).
%~\footnote{It should be noted that there is a gradual crossover between  $(\chi_C=C/T)$ and $\chi=2C/T$ as the temperature is lowered~\cite{unpub}.}. 

Eqn. \ref{Ryzhkin} contains much physics and is deserving of subtle appreciation.  The term in ${\bf H}$ represents the normal drift current of the charges $Q_\pm$ in the applied field ${\bf H}$, and if there were only this term, then spin ice would be represented as a true conductor, precisely equivalent to an electrolyte. However the term in ${\bf M}$ opposes the direct current and indeed extinguishes it completely at equilibrium, where ${\bf M} = \chi_T {\bf H}$. 
This reaction field does not originate in the magnetic monopoles themselves but rather in the configurational entropy of the monopole vacuum: magnetization of the system reduces that entropy and hence provides a thermodynamic force that opposes the current~\cite{Ryzhkin}. It should also be noted that what stops the current in Ryzhkin's formulation is not the sample boundaries: this is correct under the approximation that the system is homogenous and linear. If one further allows the competition of diffusion and drift to set up charge density gradients then the boundaries immediately become relevant and one must additionally consider boundary conditions that do not allow the passage of monopoles. However this is not necessary in the approximation considered. Finally it should be emphasised that 
the extinction of the current implied by Eqn. \ref{Ryzhkin} holds only for very small field and magnetization, for it is  only in this limit 
that Eqn. \ref{entropy} is valid. 

At first sight the right hand side of Eqn. \ref{Ryzhkin} is zero but this is only true at infinite time. 
By introducing a frequency Fourier transform of the magnetization and field and combining Eqn. \ref{flux} with Eqn. \ref{Ryzhkin}, 
Ryzhkin found that
\begin{equation}\label{acchi}
\chi(\omega) = \frac{\chi_T}{1 -i \omega\tau},
\end{equation}
where the inverse relaxation time $\tau^{-1} = \kappa/\chi_T$. Eqn. \ref{Ryzhkin} can also be integrated to predict the magnetic 
response to the sudden application or removal of a uniform field (see Section. 12). Assuming an ellipsoidal sample, when a uniform external field ${\bf H}_{ext}$ is applied, the internal field ${\bf H}_{int}$ is reduced by the demagnetizing field $\mathcal{D}{\bf M}$, such that ${\bf H}_{int} = {\bf H}_{ext}- \mathcal{D} {\bf M}$. In spin ice the demagnetizing field arises from the magnetic pole density associated with uncompensated surface monopoles. As the spin ice sample is magnetized, an imbalance of magnetic charge develops at opposite surfaces as a result of the transient monopole current described by Eqn. \ref{Ryzhkin}. However, as a result of the entropic `reaction field' discussed above, the monopoles are not sufficiently free to achieve complete screening of the internal field. The incomplete screening of 
of the internal field due to magnetic monopoles has been discussed in detail by Ryzhkin and Ryzhkin~\cite{Ryzhkin-new}.

\vspace{0.5cm}
\section{Definition of the two characteristic rates $\nu$ and $\nu_0$}

We may define the relaxation rate $\nu$ by: 
\begin{equation}\label{rate0}
\nu = \tau^{-1} = \frac{\kappa}{\chi_T} = \frac{\mu_0 u Q x}{V_0 \chi_T},
\end{equation}
where the concentration $c$ has been substituted for the total dimensionless monopole density or mole fraction $x = c V_0$, and 
$V_0$ is the volume per site of the diamond lattice inhabited by the magnetic monopoles: $V_0 = (8/3\sqrt{3}) a^3$, where
$a$ is the near neighbour spacing on the diamond lattice.  

Using the Nernst-Einstein relation, the mobility $u$ may be replaced by the diffusivity $D$:
\footnote{The Nernst-Einstein equation, which may be derived from the Boltzmann transport equation, is one of the basic equations of electrochemistry.
According to Wannier~\cite{Wannier}, it was used by Nernst in 1888 to make the first direct measurement of the elementary electronic charge - at a time when the electron had not yet been identified and even the existence of atoms or ions was controversial. The Nernst-Einstein equation may be used in this way because diffusivity and mobility are independently measurable for an electrolyte. However, for magnetic monopoles in  spin ice no way has yet been identified to measure these quantities independently.
}
\begin{equation}\label{mobility}
u =\frac{DQ}{kT},
\end{equation}
and hence
\begin{equation}
\nu
%= \frac{\mu_0 D Q^2 x}{V_0 kT \chi_T} 
= \left(\frac{D}{\chi_T}\right) l_D^{-2},
\end{equation}
where $l_D$ is the Debye length~\footnote{In the temperature range considered here (e.g. 0.4 - 10.0 K for ${\rm Dy_2Ti_2O_7}$ ), at zero applied field, the Debye length is is the dominant length scale in the system, and as shown below, magnetic inhomogeneities develop on this length scale. In the lower temperature range, not considered here, inhomogeneities on the scale of the Bjerrum length
$l_T = \mu_0Q^2/8\pi kT$
become particularly important, and probably dominate: see Refs. \cite{Nature,NatPhys, dimensional}.}:
\begin{equation}
l_D = \left(\frac{kTV_0}{\mu_0Q^2x}\right)^{1/2}.
\end{equation}

In turn $D$ is determined by the monopole hopping frequency $\nu_0$. In a simple lattice diffusion approximation we may write~\cite{NatPhys}:
\begin{equation}\label{ddd}
D = \frac{a^2 \nu_0}{6},
\end{equation}
where $a$ is the diamond lattice constant (the numerical factor of 6 may be modified very slightly when the fact that a monopole may only hop in three out of four local directions is accounted for~\cite{CMS2}). Using the definitions $Q = 2 \mu/a$ where $\mu$ is the rare earth magnetic moment, and $\chi_C = \mu_0 \mu^2/3V_0kT$, Eqns. \ref{rate0}, \ref{mobility} and \ref{ddd} may be rearranged to give:
\begin{equation}\label{x}
 \nu = g \nu_0 x,
\end{equation}
where $g = \chi_C/\chi_T = 1/2$ in Ryzhkin's theory but is more generally weakly temperature dependent and varies between $g=1$ at high temperature and $g=2$ at low temperature~\cite{TSF}.
Equation \ref{x} will be seen to be very important for the interpretation of experiments on spin ice. 

\vspace{0.5cm}
\section{Coulombic Correlation of the Monopole Current}

We define the flux of positive and negative monopoles as ${\bf j}_+$ and ${\bf j}_-$ respectively. Assuming there is no temperature gradient in the system, then the thermodynamic equations of motion are (with $i,j=+,-$):
\begin{equation}
{\bf j}_i = L_{ii} {\bf X}_i + L_{ij} {\bf X}_j, 
\end{equation}
where ${\bf X}$ denotes a generalised thermodynamic force. 
If we assume that the monopole density is small, then following the theory of weak electrolytes we would expect the cross terms $L_{ij}$ with $i\ne j$ to be zero. However the monopole motion should be strongly correlated in the sense that it always acts to maintain local charge neutrality: 
\begin{equation}\label{cur}
Q_+{\bf j}_+ + Q_-{\bf j}_- = 0,
\end{equation}
\begin{equation}
Q_+c_+ + Q_-c_- = 0.
\end{equation}
It is important to emphasise that this thermodynamic force is not the same as Ryzhkin's reaction field which is a purely spin phenomenon peculiar to spin ice. In fact, Eqn. \ref{Ryzhkin} does not account for Coulombic correlation between magnetic monopoles and in the next level of description this needs to be accounted for.  

The left hand side of Eqn. \ref{cur} is simply the magnetic diffusion current ${\bf J}_{\rm diffusion}$ which contributes to the total magnetic current ${\bf J}  = \partial {\bf M}/\partial t$. At equilibrium in zero applied magnetic field, the monopole diffusion is such that it does not change the local magnetization of the system. Thus positive and negative monopoles tend to move in the same direction.  After a perturbation, the local magnetic current relaxes to zero, even though the monopoles continue to hop around the system. 
The magnetic diffusion current is given by:
\begin{equation}
{\bf J}_{\rm diffusion} = \sum_{i=\pm} Q_i{\bf j}_i({\bf r})  = - DQ \nabla \delta c({\bf r}),
\end{equation}
where $\delta c({\bf r}) = c_+({\bf r})-c_-({\bf r})$. In the zero field equilibrium state the average local gradient of charge density is everywhere zero.

\vspace{0.5cm}
\section{Spatial Dependence of Longitudinal Magnetization}\label{Sec5}

The Coulombic correlations create a diffusion force that tends to smooth the local longitudinal magnetization. 
This may be seen as follows. By Helmholtz' theorem the vector field ${\bf M}({\bf r})$ can be decomposed into an irrotational or longitudinal ($i$) part and a solenoidal or transverse ($s$) part:
\begin{equation}
{\bf M}({\bf r}) = {\bf M}^i({\bf r}) + {\bf M}^s({\bf r}).
\end{equation}
The spin ice ground state is defined by the condition ${\bf M}^i({\bf r}) = 0$, which gives
\begin{equation}
{\bf M}^{\rm ground}({\bf r}) = {\bf M}^s({\bf r}).
\end{equation}
Physically, this is a consequence of the spin ice ground state consisting of closed loops of spins. 
The irrotational or longitudinal part is finite only as a result of thermal excitation of magnetic monopoles. As $\nabla \times{\bf M}^i({\bf r}) = 0$, the vector Laplacian is simply 
\begin{equation}\label{Lap}
\nabla^2 {\bf M}^i({\bf r}) =\nabla(\nabla \cdot {\bf M}^i({\bf r})), 
\end{equation}
a result that will be used below. Henceforth (unless otherwise stated) we shall only deal with the longitudinal magnetization and the superscript $i$ will be dropped. 

Defining $\phi({\bf r})$ as the magnetic scalar potential, the local magnetic field is: 
\begin{equation}
{\bf H}({\bf r}) = -\nabla \phi({\bf r}).
\end{equation}
and by Poisson's equation and Maxwell's equation, the local magnetic charge density is: 
\begin{equation}
Q\delta c({\bf r}) =  Q(c_+({\bf r})- c_-({\bf r})) = \nabla\cdot{\bf H}({\bf r}) = -\nabla^2 \phi({\bf r}) = -\nabla \cdot {\bf M}({\bf r}), 
\end{equation}
Thus, using Eqn. \ref{Lap}, the local charge density gradient is: 
\begin{equation}
Q\nabla \delta c{\bf (r)} = -\nabla^2 {\bf M}({\bf r}).
\end{equation}
The magnetic diffusion current associated with a finite charge density gradient is: 
\begin{equation}
{\bf J}_{\rm diffusion} = -D Q\nabla \delta c{\bf (r)} = D\nabla^2 {\bf M}({\bf r}).
\end{equation}
This term can then be added to Eqn. \ref{Ryzhkin} to describe relaxation of the spatial charge arrangements: 
\begin{equation}\label{newcurrent}
 {\bf J}({\bf r})  = \kappa \left({\bf H}({\bf r}) - \chi_T^{-1}{\bf M}({\bf r})\right) + D \nabla^2 {\bf M({\bf r})}
\end{equation}
In recent work Ryzhkin and Ryzhkin~\cite{Ryzhkin-new} stated such an equation to facilitate a calculation of 
magnetic screening effects in spin ice. 

\vspace{0.5cm}
\section{Free Energy Functional}

The same equation can be derived from a Landau-type free energy functional as follows. If we apply a local field ${\bf H}({\bf r})$ 
then this induces a longitudinal response ${\bf M}({\bf r})$. The local magnetization is opposed by the entropy cost of ordering the spins of the sample as well as the entropy cost of creating a local charge imbalance. Note that these two factors are distinct: it is possible to increase order in the sample without creating a local charge imbalance. 

From general chemical thermodynamics we expect the entropic cost of charge imbalance to make the following contribution to the local Gibbs free energy: 
 \begin{equation}
G'_{\rm local} = \frac{kT}{2} \frac{(\delta c({\bf r}))^2}{c}.
\end{equation}
Hence using Poisson's equation and Maxwell's equation again:
\begin{equation}
G'_{\rm local} = \frac{kT}{2} \frac{\left(\nabla\cdot {\bf M}\right)^2}{Q^2c_0} =  \frac{\mu_0}{2} l_D^2 \left(\nabla\cdot {\bf M}\right)^2
= \frac{\mu_0D}{2\kappa} \left(\nabla\cdot {\bf M}\right)^2
\end{equation}
We may then write down a free energy functional for the system:
\begin{equation}\label{Landau}
G({\bf M}({\bf r})) = \int \frac{\mu_0}{2\chi_T}M({\bf r})^2 - \mu_0 {\bf M}({\bf r})\cdot{\bf H}({\bf r}) + \frac{\mu_0D}{2\kappa} \left(\nabla\cdot {\bf M}\right)^2 d {\bf r},
\end{equation}
where the first term on the right represents the Jaccard entropy~\cite{Ryzhkin} which in this representation is seen to be equal to the
entropy of a cooperative paramagnet~\cite{Henley}. 
The rate of change of longitudinal magnetization ${\bf M}({\bf r})$ may be found in a linear response approximation:
\begin{equation}\label{func}
\frac{\partial {\bf M} ({\bf r})}{\partial t} = -\frac{\kappa}{\mu_0}\left[\frac{\delta G({\bf M}({\bf r}))}{\delta {\bf M}({\bf r})} \right]
= \kappa \left[{\bf H}({\bf r})-\chi_T^{-1} {\bf M}({\bf r})\right] + D \nabla^2 {\bf M({\bf r})}.
\end{equation}
which gives Eqn. \ref{newcurrent}. The derivation of the right hand term of Eqn. \ref{func} is given in the footnote
\footnote{
The functional derivative of $\tilde{G}=\int (\nabla\cdot {\bf M})^2 d{\bf r}$ is found as follows:
\begin{equation}
\left\langle \frac{\delta \tilde{G}[{\bf M}({\bf r})]}{\delta {\bf M}}, {\bf f}\right\rangle
=\frac{d}{d \epsilon} 
\int \{\nabla \cdot ({\bf M}+ \epsilon {\bf f})\}^2 d{\bf r} ~~~|_{\epsilon=0}=\int  2(\nabla\cdot{\bf f})(\nabla \cdot {\bf M}) d{\bf r},
\end{equation}
where the angular brackets indicate a distribution with vector test function ${\bf f}({\bf r})$.
Now 
\begin{equation}
\nabla\cdot[({\bf f})(\nabla \cdot {\bf M})]=(\nabla\cdot{\bf f})(\nabla \cdot {\bf M})+({\bf f}) (\nabla (\nabla\cdot{\bf M}) ),
\end{equation}
%and, using Eqn. \ref{Lap},
%\begin{equation}
%\nabla\cdot[({\bf f})(\nabla \cdot {\bf M})]=(\nabla\cdot{\bf f})(\nabla \cdot {\bf M})+({\bf f}) (\nabla^2{\bf M}),
%\end{equation}
%where the fact that grad div equals the vector Laplacian follow from the fact that the field ${\bf M}$ is irrotational.   
%\begin{equation}
%(\nabla\cdot{\bf f})(\nabla \cdot {\bf M}) =\nabla\cdot[({\bf f})(\nabla \cdot {\bf M})]- ({\bf f}) (\nabla^2{\bf M}) 
%\end{equation}
so using Eqn. \ref{Lap} and the divergence theorem, 
%\begin{equation}
%\left\langle \frac{\delta \tilde{G}[{\bf M}({\bf r})]}{\delta {\bf M}}, {\bf f}\right\rangle
%=2 \int \nabla\cdot[({\bf f})(\nabla \cdot {\bf M})]- ({\bf f}) (\nabla^2{\bf M}) 
%~d{\bf r}
%\end{equation}
\begin{equation}
\left\langle \frac{\delta \tilde{G}[{\bf M}({\bf r})]}{\delta {\bf M}}, {\bf f}\right\rangle
=2 \int_S [({\bf f})(\nabla \cdot {\bf M})]\cdot{\bf n} da - \int 2 ({\bf f}) (\nabla^2{\bf M}) 
~d{\bf r},
\end{equation}
where $S$ denotes a surface integral, ${\bf n}$ a unit normal to the surface and $a$ a surface element. If the surface charge is everywhere zero then we have:
\begin{equation}
\left\langle \frac{\delta \tilde{G}[{\bf M}({\bf r})]}{\delta {\bf M}}, {\bf f}\right\rangle
= - \int 2 ({\bf f}) (\nabla^2{\bf M}) 
~d{\bf r},
\end{equation}
and finally
\begin{equation}
 \frac{\delta \tilde{G}[{\bf M}({\bf r})]}{\delta {\bf M}}
= - 2 (\nabla^2{\bf M}).
\end{equation}
}, from which it can be seen that the contribution of surface charge to the Gibbs energy is neglected.

Under conditions of fixed temperature and field, the rate of dissipation is:
\begin{equation}
 \left(T \frac{\partial S}{\partial t}\right)_{\rm irreversible} =  -\left( \frac{\partial G}{\partial t}\right) = 
- \frac{\partial {\bf M}({\bf r}, t)}{\partial t} \frac{\delta G({\bf r}, t)}{\delta {\bf M}}.
\end{equation}
Hence, using Eqn. \ref{flux}, 
\begin{equation}\label{func}
 \left(T \frac{\partial S}{\partial t}\right)_{\rm irreversible} = \mu_0
{\bf J}\cdot
 \left[ {\bf H}({\bf r})-\chi_T^{-1} {\bf M}({\bf r})+ \frac{D}{\kappa} \nabla^2 {\bf M({\bf r})}\right],
\end{equation}
which is the extension of Eqn. \ref{entropy} to include monopole diffusion. 

Owing to the neglect of surface charge, these equations are generally applicable only under conditions of small field and small magnetization, or else at short time. When these conditions are violated the build up of surface charge may have a decisive influence on the internal fields and on the magnetization process, and the preceding equations would need to be modified to account for this.  

\vspace{0.5cm}
\section{Generalised Longitudinal Susceptibility}

We introduce the spatial and time dependent Fourier transforms of the longitudinal magnetization and longitudinal magnetic field:
\begin{equation}
{\bf M}({\bf r},t) = \frac{1}{2\pi V} \sum_{\bf q} \int {\bf M}({\bf q,\omega})e^{i{\bf q}\cdot{\bf r}-i\omega t} d\omega,
\end{equation}
\begin{equation}
{\bf H}({\bf r},t) = \frac{1}{2\pi V} \sum_{\bf q} \int  {\bf H}({\bf q,\omega})e^{i{\bf q}\cdot{\bf r}-i\omega t} d\omega,
\end{equation}
and the generalised susceptibility (assuming translational invariance):
\begin{equation}
\chi({\bf q}, \omega) = \frac{{\bf M}({\bf q},\omega) }{{\bf H}({\bf q},\omega) }.
\end{equation}
Substituting these definitions into Eqn. \ref{newcurrent} and Eqn. \ref{flux}, we find:
%\begin{equation}
 %-i \omega {\bf M}({\bf q}, \omega) =   \kappa ({\bf H({\bf q}, \omega)} - \chi_T^{-1}{\bf M}({\bf q}, \omega)) - D q^2 {\bf M}({\bf q}, \omega).
%\end{equation}
%\begin{equation}
 %-i \omega \chi({\bf q}, \omega) {\bf H}({\bf q}, \omega) =   \kappa ({\bf H({\bf q},\omega)} - \chi({\bf q}, \omega)\chi_T^{-1}{\bf H}({\bf q},\omega))
 %- D q^2 \chi({\bf q}, \omega){\bf H}({\bf q}, \omega)..
%\end{equation}
%\begin{equation}
 %-i \omega \chi({\bf q}, \omega)  =   \kappa -\chi({\bf q}, \omega)(\kappa \chi_T^{-1}+ Dq^2),
%\end{equation}
\begin{equation}\label{cqo}
\chi({\bf q}, \omega) = \frac{\kappa}{\nu_{\bf q} -i \omega}.
\end{equation}
where:
\begin{equation}\label{nuq}
\nu_{\bf q} = D q^2 + \nu.
\end{equation}
Note that a generalised susceptibility of this sort could also be derived by solving a Langevin equation incorporating the Landau free energy, as described in Ref. \cite{Goldenfeld}.  

Using Eqn. \ref{x}, the generalised susceptibility can also be written:
\begin{equation}\label{chiqw}
\chi({\bf q},\omega) = \frac{\chi_T}{1+(a^2q^2/6gx)- i \omega \tau} =   \frac{\xi^{-2} \chi_T}{\xi^{-2}+q^2- 6g i \omega \tau_0} 
%= \frac{1}{\chi_T^{-1}+ q^2l_D^2/6 -i\omega \kappa^{-1}},
\end{equation}
where $\tau_0 = 1/\nu_0$ and the correlation length is: 
\begin{equation}\label{xi}
\xi = \frac{a}{\sqrt{6gx}}.
\end{equation}

It should be emphasised that this is an equation for the longitudinal susceptibility only (${\bf M} \parallel {\bf H} \parallel {\bf q}$).
Whether at equilibrium or not, the latter is finite only if there is a finite density of monopoles. In contrast (see Section \ref{Sec5}) the transverse susceptibility of spin ice could take a paramagnetic value at equilibrium even in the complete absence of monopoles, as it does in the monopole-free spin ice ground state. In principle the transverse susceptibility could relax through the flipping of closed loops of spins~\cite{BramwellHarris}, though in reality it is more likely that it relaxes through the transient passage of magnetic monopoles. A more complete description of the wavevector dependence of the susceptibility is given in Refs. \cite{YA, Henley, Henley-C, Fennell}, and the possibility of quantum mechanical effects giving rise to transverse dynamics distinct from magnetic monopoles (so called `photons') has been discussed in Refs. ~\cite{Shannon,Benton}.

There is potentially a problem with Eqn. \ref{nuq} and the subsequent equations. To see this we rewrite $\nu_{\bf q}$ as follows:
\begin{equation}
\nu_{\bf q} = \nu_0\left(\frac{q^2a^2}{6}+ gx\right),
\end{equation}
and assume that the maximum possible equilibrium value of the density is $x = 1/2$. Since it seems implausible that $\nu_{\bf q}$ would ever exceed $\nu_0$ (and indeed $\nu_0$ should generally be less than it), then it appears that Eqn. \ref{nuq} breaks down at $q^2a^2 > 3 g$. To guarantee this never happens we can write:
\begin{equation}
\nu'_{\bf q} = \nu_0\left(\frac{q^2a^2}{6(1+ A q^2a^2/3)}+ gx\right),
\end{equation}
where $A$ is introduced as a phenomenological (dimensionless) parameter. Applying Eqn. \ref{cqo} we find: 
\begin{equation}\label{chinew}
\chi({\bf q},\omega) = \frac{\xi^{-2} \chi_T}{\xi^{-2}+q^2/(1+ A q^2a^2/3)- 6g i \omega \tau_0}.
\end{equation}
As discussed further below (Section \ref{Neutron}), it would seem more realistic to use this phenomenological equation than Eqn. \ref{chiqw} in order to describe experiment.  

\vspace{0.5cm}
\section{Equilibrium Field Fluctuations due to Monopoles}

Consider now the spin ice state in zero applied magnetic field, where the internal field ${\bf H}({\bf r})$, which originates from the magnetic monopoles, is no longer constrained but instead relaxes self consistently with the magnetization. An approximation to the problem is that close to equilibrium, the field takes the value ${\bf H}({\bf r}) = \chi_T^{-1} {\bf M}({\bf r})$ everywhere, and that this relation is maintained for small fluctuations away from equilibrium. 
The internal field therefore costs spin entropy: 
\begin{equation}
-TS_{\rm field} = \mu_0\frac{\chi_T}{2} {H}({\bf r})^2,
\end{equation}
and energy 
\begin{equation}
U_{\rm field} = \frac{1}{2} {\bf B}\cdot{\bf H} = .\frac{\mu_0}{2}H^2(1+ \chi_T),
\end{equation}
but this is offset by the energy gain in magnetizing the sample:  
\begin{equation}
U' = -\mu_0 {\bf M}\cdot{\bf H} = -\mu_0\chi_TH^2.
\end{equation}
Summing these contributions and allowing the magnetic charge density to fluctuate along with the field we find a free energy functional for field fluctuations: 
\begin{equation}\label{Gfield}
F({\bf H}({\bf r})) = \int \frac{\mu_0}{2}H({\bf r})^2 + \frac{\mu_0D}{2\kappa} \left(\nabla\cdot {\bf H}\right)^2 d {\bf r}.
\end{equation}
This functional is entirely equivalent to that for electric field fluctuations in an electrolyte and in fact is equivalent to Debye-H\"uckel 
theory~\cite{Oosawa}. Thus we see that by suppressing spin fluctuations (i.e. setting ${\bf H}({\bf r}) = \chi_T^{-1} {\bf M}({\bf r})$) we recover the Debye-H\"uckel approximation of Castelnovo {\it et al.}~\cite{CMS2}.  

Introducing the Fourier transformed field ${\bf H}({\bf r}) = V^{-1} \sum_{\bf q} {\bf H}({\bf q})e^{i {\bf q} \cdot {\bf r}}$ and substituting into Eqn. \ref{Gfield} we find:
%\begin{equation}
%F({\bf H}) = \int \frac{\mu_0}{2}\left(\sum_{\bf q}{\bf H}({\bf q})e^{i {\bf q}\cdot{\bf r}}\right)^2 
%+ \frac{\mu_0D}{2\kappa} \left(\sum_{\bf q}{\bf q}\cdot {\bf H}({\bf q})e^{i {\bf q}\cdot{\bf r}}\right)^2  d {\bf r}.
%\end{equation}
%\begin{equation}
%F({\bf H}) = \int \frac{\mu_0}{2}\left(\sum_{\bf q} \sum_{\bf q'} {\bf H} ({\bf q}) \cdot {\bf H}({\bf q'})
%+ \frac{\mu_0D}{2\kappa}({\bf q}\cdot  {\bf H} ({\bf q})) ({\bf q}'\cdot {\bf H}({\bf q'})\right)e^{i ({\bf q+q}')\cdot{\bf r}}   d {\bf r}.
%\end{equation}
\begin{equation}
F = \frac{\mu_0V}{2}\sum_{\bf q}  { H} ({\bf q})^2  (1+ q^2 l_D^2),
\end{equation}
where $V$ is the volume and we have used $D/\kappa= l_D^2$ and the fact that the field, being derived from a scalar potential, is longitudinal to the wavevector. Since the probability of a fluctuation is $\propto e^{-F/kT}$, we immediately see that
the mean square amplitude of a mode is: 
\begin{equation}
\langle ({H}_{\bf q})^2\rangle = \frac{kT}{(1+q^2 l_D^2)\mu_0V}.
\end{equation}
The mean square field at a point is:
\begin{equation}\label{mean}
\langle H({\bf r})^2\rangle = \sum_{\bf q} \langle ({H}_{\bf q})^2\rangle 
\approx \frac{V}{(2\pi)^3}\int_0^{\pi/a} \frac{4\pi q^2kT}{(1+q^2 l_D^2)\mu_0V} dq.
\end{equation}
The integral is dominated by short wavelength modes and the mean square field approximately takes the value: 
\begin{equation}\label{meansq}
\langle H({\bf r})^2\rangle
\approx \frac{kT}{2\pi l_D^2 a\mu_0}.
\end{equation}

\section{Mean square induction at a point}

Using Eqn. \ref{meansq}, the mean square induction at a point is: 
\begin{equation}
\langle B^2 \rangle = \frac{\mu_0 kT}{2\pi l_D^2 a}(1+ \chi_T)^2.
\end{equation}
Obviously this equation depends on the induction being averaged over a sufficiently large volume that contains locally magnetized spins. If the point with which we are concerned experiences no local induction from the magnetized spins, and sees only a far field, then the mean square induction is simply  
\begin{equation}\label{far}
\langle B^2 \rangle_{\rm far} = \frac{\mu_0 kT}{2\pi l_D^2 a}.
\end{equation}
Recalling that $l_D^{-2} = \mu_0 Q^2 x/ kT V_0$ where $V_0 \sim a^3$ it may be shown that
\begin{equation}\label{near}
\langle B^2 \rangle_{\rm far} \approx B_0^2 x 
\end{equation}
where 
\begin{equation}
B_0 \sim \frac{\mu_0 Q}{a^2},  
\end{equation}
and the symbol $\sim$ is used to indicate that factors of order unity are dropped. In the case that there is local induction arising from magnetized spins, this equation may be modified to: 
\begin{equation}
B_0' \sim \frac{\mu_0 Q}{a^2}(1+\chi_T). 
\end{equation} 

It is instructive to derive Eqn. \ref{near} in direct space. At a point in the sample
the squared field may be averaged over contributions at distance $r$ weighted by the probability of finding a monopole at that distance: 
\begin{equation}
\langle B^2\rangle \approx \sum \left(\frac{\mu_0 Q}{4\pi r^2}\right)^2 \left(\frac{4\pi r^2 x}{a^2}\right) 
\sim \frac{B_0^2 x}{a}  \int_a^{\infty} \frac{a^2}{r^2} dr =  B_0^2 x,
\end{equation}
which neglects correlation between the field contributions. The fields are correlated over a distance of $l_D$, but the average number of monopoles within a distance $l_D$ is typically of order unity, so correlation may be neglected to a first approximation.  
The field $B_0$ is approximately that due to a monopole or a spin at a distance $a$. Thus if a defect is viewed at a distance $a$ it looks like a spin, but if viewed at a much greater distance it looks like a monopole. For this reason monopoles are best detected by measuring their far fields~\cite{Nature}. 

\vspace{0.5cm}
\section{Relaxation of the Field fluctuations}

The equivalent electrolyte theory has been formulated and worked out in detail by Oosawa~\cite{Oosawa}, who
found that the relaxation rate of a mode labelled by ${\bf q}$ is: 
\begin{equation}\label{rate1}
\nu_{\bf q} = D(q^2 + l_D^{-2}) = \nu_0\left(\frac{a^2q^2}{6}\right)+ \kappa. 
\end{equation}
Thus, short wavelength modes relax at a rate of $\nu_0$, the monopole hop rate, and long wavelength modes relax at a rate $\kappa$, the monopole conductivity.  Fluctuations are important on all scales between the lattice constant and the Debye length, so there is a dispersion of relaxation rates from the monopole hop rate $\nu_0$ to the bulk field relaxation rate, or monopole conductivity $\kappa$. 

From Eqn. \ref{mean}, it may be seen that monopoles at distance $a$ and those at distance $l_D$ make similar contributions to the mean square field, while those at much greater distance may be neglected (in zero applied field). Monopoles at distance $a$ reverse the local field at a rate of approximately $\nu_0$, whereas the cloud of monopoles at distance $l_D$ only reverses the field at the much slower rate $\kappa$, although it gives rise to small field fluctuations at a rate $\nu_0$. 

Comparison of Eqn. \ref{rate1} with Eqn. \ref{rate0} suggests that the field correlations relax at a rate $\kappa$ while the spin correlations relax at a rate 
$\nu = \kappa/\chi_T$. Although this difference may reflect the approximations made, it also seems plausible that the spin correlations relax more slowly than the field correlations at low temperature. Thus the spin system can only relax by the passage of monopoles, so the time taken to find the most probable spin arrangement will generally be longer than the time taken to find the most probable monopole arrangement. 

\vspace{0.5cm}
\section{Spin Lattice Relaxation}

It is obvious that the relaxation rate $\nu$ is equal to the spin-lattice relaxation rate $\tau_1^{-1}$ but it is useful to see how this arises in detail. 
Initial application of a magnetic field $H$ should result in almost instantaneous magnetization, in which energy is stored within the system of effective spins. The spin temperature $T_s$ is therefore initially higher than the applied bath temperature $T$. Following Casimir and du Pr\'e~\cite{Casimir,Morrish} the temperature difference $\theta = T_s-T$ determines the rate of exchange of heat $\mathcal{Q}$ with the thermal bath, and the consequent return of the spin system to thermal equilibrium at temperature $T$: 
\begin{equation}\label{alph}
\frac{d\mathcal{Q}}{dt} = -\alpha \theta.
\end{equation}
We may also write:
\begin{equation}
d\mathcal{Q} = C_H dT_s
\end{equation}
where $C_H$ is the heat capacity at constant field. Hence we find: 
\begin{equation}
\frac{d T_s}{dt} = -\frac{\alpha}{C_H} \left(T_s - T\right),
\end{equation}
and the spin temperature relaxes at a rate $\tau_1^{-1} = \alpha/C_H$. 

To link this to the magnetization relaxation we use the thermodynamic relations
\begin{equation}
d\mathcal{Q} = TdS = C_MdT-\mu_0 T\left(\frac{\partial H}{\partial T}\right)_M \chi dH,
\end{equation}
and
\begin{equation}
\chi dH = \chi_T dH +  \left(\frac{\partial M}{\partial T}\right)_H dT,
\end{equation}
where $\chi dH = dM$ and $C_M$ is the heat capacity at constant magnetization. If then we apply a field $H = H_1e^{i\omega t}$ and elicit a response $\theta = \theta_1 e^{i\omega t}$ the above two equations may be solved with the substitutions $dH \rightarrow H, dT \rightarrow \theta$, $d\mathcal{Q} \rightarrow -\alpha \theta dT$ (Eqn. \ref{alph}) and use of the thermodynamic relation $C_H-C_M = -\mu_0 T (\partial H/\partial T)_M (\partial M/\partial T)_H$. 
This gives the well-known result~\cite{Morrish}: 
\begin{equation}
\chi = \left[\chi_S+ \frac{\chi_T-\chi_S}{1+\omega^2\tau_1^2}\right] + i \left[ \frac{(\chi_T-\chi_S)\omega \tau_1}{1+ \omega^2\tau_1^2}\right],
\end{equation}
where $\chi_S = \chi_T C_M/C_H$ is the adiabatic susceptibility. By comparison with Eqn.  \ref{acchi} we see that in Ryzhkin's approximation $\nu = \tau_1^{-1}$ and the adiabatic susceptibility is assumed to be zero. The former is easily understood as any magnetization involves the passage of a monopole current accompanied by dissipation. However, it is conceivable that the adiabatic response could be finite in the real material, and involve the `stretching' of the excited state magnetic moment along the field direction, in which case the adiabatic susceptibility $\chi_S$ could be a direct measure the density of excited states or monopoles. This idea needs to be checked in detail. 
 
 \subsection{Phonon Bottleneck}
 
We may also modify this approach to include a `phonon bottleneck'. The spin system is considered to be connected to the phonon system at temperature $T_p$, and the phonon system is connected to the bath at temperature $T$. The thermal relaxation between phonon system and bath is characterised by a thermal conductivity $\alpha'$. If we make a steady state approximation to the 
phonon temperature, then the rate of heat exchange between phonon system and bath is simply: 
\begin{equation}
\frac{d\mathcal{Q}}{dt} = \frac{dE}{dt} = -\alpha(T_p-T)
\end{equation}
where $E = U-\mu_0 M H$ is the magnetic enthalpy.  Under the approximation that the monopole internal energy $U$ is constant, we find simply that:
\begin{equation}\label{bottle}
T_p = T + (\alpha')^{-1} \mu_0 H \left(\frac{dM}{dt}\right). 
\end{equation}
Thus a thermometer placed on the sample could be used to measure $T_p$ and hence gain an alternative measure of the magnetic current 
$dM/dt$ after transients have died away.  

The rise in temperature of the sample (Eqn. \ref{bottle}) occurs when the rate of flow of heat between the spin and phonon systems exceeds the rate of flow of heat from the phonon system to the bath. In the steady state approximation the criterion for this is $\alpha' < \alpha = \nu C_H$. 
In the low temperature limit we find $C_H \approx (|\mu|/kT^2V_0)x$~\cite{Zhou}, where $\mu$ is the monopole chemical potential~\cite{CMS,Jaubert} and hence: 
\begin{equation}
\frac{x}{T} > \sqrt{\frac{\alpha' k}{|\mu| g\nu_0}},
\end{equation} 
is a criterion for observation of this effect. The ratio $x(T, H = 0)/T$ is always sufficiently small that this analysis suggests that the bottleneck can never be observed in zero applied field, and hence any observation of a bottleneck is likely to reflect a significant field-induced increase in $x(T, H)$ (the Wien effect). This conclusion is consistent with the experimental observations of Slobinsky {\it et al.}~\cite{Slobinsky} 
who observed a phonon bottleneck, albeit in fields much stronger than those appropriate to the theory discussed here.

\subsection{Thermal Quench}

If bound monopole pairs equilibrate sufficiently quickly with the monopole vacuum, then the magnetic monopoles may be regarded as in direct equilibrium with the vacuum: $(0) = (+) + (-)$. The equiiibrium constant is 
\begin{equation}
K_{eq} = x_0^2 
\end{equation}
where we temporarily label the equilibrium density as $x_0$. By definition the thermodynamic equilibrium constant is given by
$K_{eq}  = e^{2 \mu/T}$ where $\mu < 0$ is the monopole chemical potential~\cite{CMS,Jaubert}. 

\begin{figure}[h]
\includegraphics[width=0.9\linewidth]{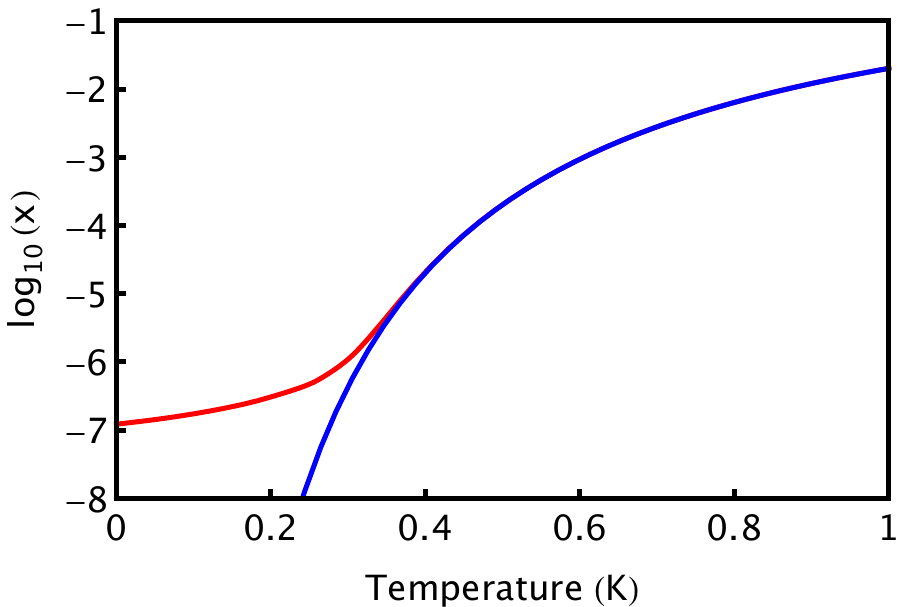}
\caption{
\label{fig1} 
Thermal evolution of the monopole density $x$ according to Eqn. \ref{cool} (red line) versus the equilibrium density $2\exp(\mu/T)$ (blue line), with the cooling process starting at 1.0 K at a rate $10^{-5}$ K s$^{-1}$ (here $\nu_0 = 10^3$ s$^{-1}$, $\mu = -4.6$ K). }
\end{figure}

\begin{figure}
\includegraphics[width=0.9\linewidth]{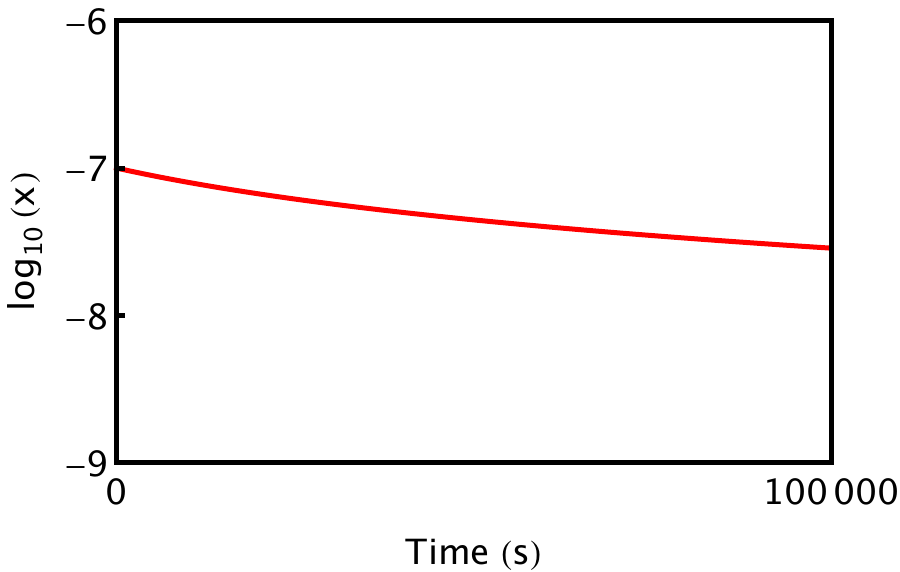}
\caption{
\label{fig2} 
Temporal evolution of the monopole density $x$ starting at $x=10^{-7}$, according to Eqn. \ref{kin} (parameters as in Fig. \ref{fig1}).} 
\end{figure}

Neglecting Bjerrum pairs~\cite{NatPhys}, the kinetic rate equation for the change in monopole density is: 
\begin{equation}\label{kin}
\frac{dx}{d t} = - \nu_0 \left(\frac{x^2}{4} - e^{2\mu/T}\right).
\end{equation}
where the first term on the right accounts for monopole recombination and the second for monopole generation. The recombination rate constant has been assumed to be equal to the monopole hop rate. If the temperature is lowered at a rate $dT/dt = -r$ then it follows that
\begin{equation}\label{cool}
\frac{dx}{dT} = \nu_0 r^{-1} \left(\frac{x^2}{4} - e^{2\mu/T}\right).
\end{equation} 
Numerical solution of this equation shows that monopole density reaches a finite approximate steady state of the order $x=r/\nu_0$ at low temperatures. Putting in reasonable parameters for ${\rm Dy_2Ti_2O_7}$ spin ice (e.g. $\nu_0 =1000~{\rm s}^{-1}, \mu = -4.6$ K), a rate of cooling of $r = 10^{-5}$ K 
s$^{-1}$ (about 1 K per day) would result in a residual density of about $x = 10^{-7}$ at temperatures lower that 0.25 K (see Fig. \ref{fig1}). Arrest of the cooling at a base temperature significantly less than 0.25 K, where $\exp(2\mu/T)$ becomes entirely negligible, then results in a very slow power law decay of the monopole density according to (see Eqn. \ref{kin}) :
\begin{equation}\label{kin}
x(t) = \frac{x(0)}{1+ x(0)\nu_0 t/4}, 
\end{equation}
and even one day of waiting would barely reduce the density by a further power of 10 (Fig. \ref{fig2}). Therefore, with any realisable rate of cooling and time of waiting it is not possible to completely rid the system of monopoles on experimental time scales. 

The preceding analysis neglects many factors that may become important at low temperatures, including possible thermal evolution of the hop rate, extrinsic factors and kinetic constraints arising from the Dirac strings. However, most of these factors will tend to tend to reduce, rather than increase, the rate of relaxation, so it is safe to conclude that the analysis is correct in its conclusion that 
a monopole-free state in zero applied magnetic field remains inaccessible to experiment
\footnote{One may contrast the case of pure water at room temperature, where a density of $\sim10^{-9}$ H$^{+}$ ions per water molecule is easily maintained at equiliibrium on experimental time scales: however the diffusion constant of H$^{+}$ in water is some $10^8$ times larger than that of magnetic monopoles in spin ice.}. A detailed analysis of idealised thermal quenches in spin ice~\cite{Quench} has identified the important role of monopole-antimonopole pairs that cannot immediately annihilate by a single spin flip. These `noncontractable' pairs form long lived metastable states at low temperature. 
 
It is also pointed out in Ref. \cite{Quench} that at sufficiently low temperatures, owing to the divergence of the mobility as $1/T$ (Eqn, \ref{mobility}), monopoles will recombine at the maximum speed allowed by the monopole hop rate. However, in the electrolyte theory, this effect is accounted for by the concept of the Bjerrum pair~\cite{Nature, NatPhys}. Such nonlinear response occurs only within the pair, that is when the monopole-monopole separation is less than $l_T = \mu_0Q^2/8\pi kT$. This fast nonlinear response then appears like the flipping of giant dipoles of magnetic moment $Ql_T$~\cite{NatPhys}, but for monopoles at greater separation, the ordinary recombination kinetics of Eqn. \ref{kin} are obeyed. The average monopole separation grows with decreasing temperature much faster than the Bjerrum pair radius, so in a `slow' quench of the sort described above, the divergence of the mobility should not significantly speed up the rate of recombination. The role of Bjerrum pairs, which is closely connected to the Wien effect~\cite{Nature,NatPhys} is not considered further here. 
  
\vspace{0.5cm}
\section{Application to Experiment: General}

In the following sections I discuss the application of these ideas to different experimental measurements. 
The equations quoted and derived here should be applicable at sufficiently small applied field and at temperatures ($> 0.4$ K for 
Dy$_2$Ti$_2$O$_7$) where the Wien effect is absent, so the dimensionless monopole density $x$ depends only on temperature. 

\begin{figure}[h]
\includegraphics[width=0.9\linewidth]{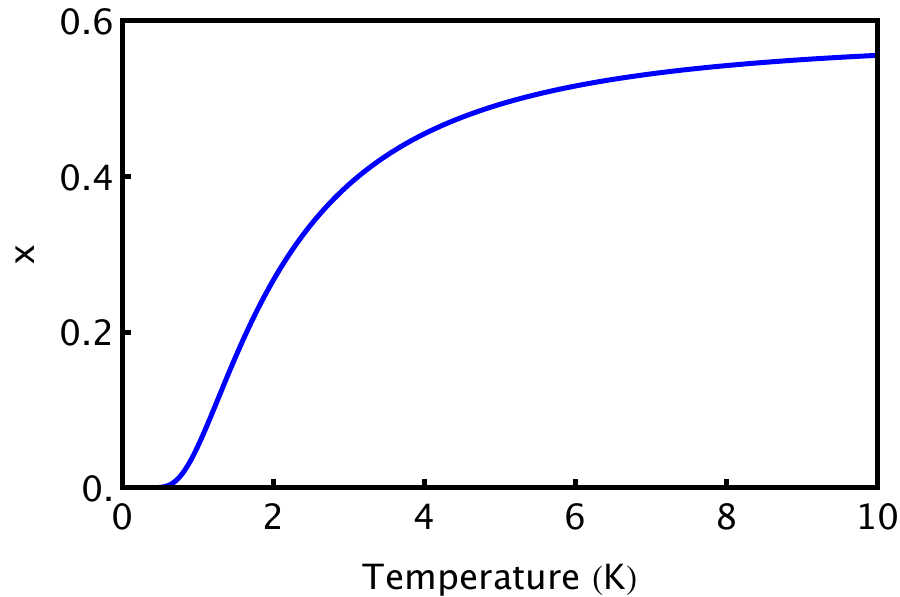}
\caption{\label{fig3} 
Approximate evolution of the dimensionless monopole density $x$ with temperature~\cite{Bloxsom} for parameters appropriate to 
Dy$_2$Ti$_2$O$_7$. The values of $x$ have been inferred from fitting experimental specific heat data to Debye-H\"uckel theory, 
according to the method of Ref. \cite{Zhou}.}
\end{figure}

It is important to emphasise that the experimental response in all cases depends on $x(T) = \nu/g\nu_0$. 
There is a general belief that the monopole hop rate $\nu_0$ is temperature independent~\cite{Jaubert}. Assuming this, $x(T)$ can be calculated by numerical simulation~\cite{Jaubert}, by Debye-H\"uckel theory~\cite{CMS2}, or approximately inferred from the specific heat~\cite{Zhou}. For 
Dy$_2$Ti$_2$O$_7$ the monopole density is roughly constant below $10$ K and decreases rapidly as the temperature is lowered below 2 K
(see Fig. \ref{fig3}). The corresponding relaxation time therefore shows a plateau between $10$ K and $2$ K, and increases rapidly as the temperature is further lowered~\cite{Jaubert}. The picture of monopoles hopping at a constant rate $\nu_0$  breaks down at temperatures above about 10 K, where an Orbach-like relaxation process involving an excited crystal field level becomes important~\cite{Ehlers1}. 

\vspace{0.5cm}
\section{Magnetization Measurements}

\subsection{dc-Magnetization}

To treat a bulk magnetization measurement we can set $q = 0$ in the above equations. In any real sample demagnetizing fields need to be accounted for.  
If we assume an ellipsoidal sample and write ${\bf H}_{\rm int} = {\bf H}_{\rm ext} - \mathcal{D} {\bf M}$ where $\mathcal D$ is the demagnetizing factor, then Ryzhkin's equation becomes: 
\begin{equation}
\frac{\partial{\bf M}}{\partial t} =  \kappa ({\bf H_{\rm ext}} - {\bf M}(\chi_T^{-1} + \mathcal{D})).
\end{equation}
For the case of a steady field this equation may be integrated to find: 
\begin{equation}
{\bf M}(t) =   \frac{{\bf H_{\rm ext}}(1-e^{-t/\tau})}{\chi_T^{-1} + \mathcal{D}} + {\bf M}(0)e^{-t/\tau},
\end{equation}
so the relaxation of the magnetization is purely exponential.
It may be seen that the susceptibility $\chi_T$ behaves as an effective demagnetizing field and that the apparent susceptibility is 
\begin{equation}
\chi_a \equiv \frac{M}{H_{\rm ext}} =  \frac{1}{\mathcal{D}+\chi_T^{-1}},
\end{equation}
which tends towards $1/\mathcal{D}$ as $T \rightarrow 0$. 

\begin{figure}[h]
\includegraphics[width=0.9\linewidth]{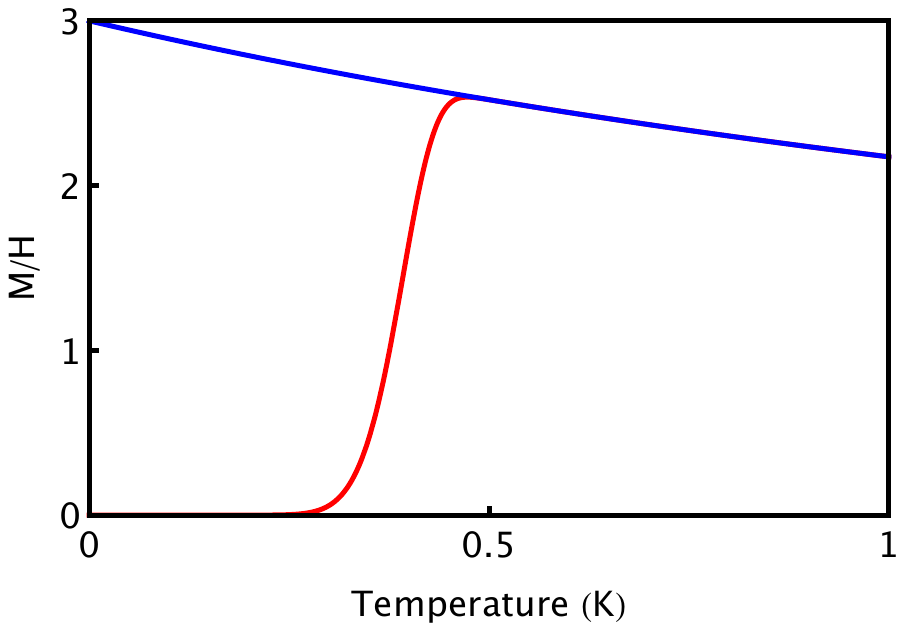}
\caption{
\label{fig4} 
Field cooled (FC, blue) versus zero field cooled (ZFC, red) splitting according to Eqn. \ref{eqzfc}, \ref{fc}, based on Ryzhkin's theory of monopole current~\cite{Ryzhkin}. 
The observation time has been set at $t_{obs} = 100$ s, 
the monopole hop rate at $\nu_0 = 10^3$ s$^{-1}$ and the demagnetizing factor at $\mathcal{D} = 1/3$.
The monopole density has been roughly approximated by 
$x \approx 2 e^{\mu/T}/\left(1+ 2 e^{\mu/T} \right)$ with $\mu = -4.6$ K,
appropriate to ${\rm Dy_2Ti_2O_7}$. 
}
\end{figure}

This equation may be used to describe field cooled (FC) and zero field cooled (ZFC) magnetization measurements. It is assumed that in the FC experiment, the sample is cooled sufficiently slowly that it always remains in equilibrium (although we have shown that this cannot be strictly true), but that in the ZFC experiment it is heated at a sufficient rate to be observed on a timescale $t_{obs} \ll \nu, \kappa$. With these approximations the FC and ZFC magnetizations are: 
\begin{equation}\label{eqzfc}
{\bf M}_{ZFC} =   \frac{{\bf H_{\rm ext}}(1-e^{-t_{obs}/\tau})}{\chi_T^{-1} + \mathcal{D}},
\end{equation}
\begin{equation}\label{fc}
{\bf M}_{FC} =  \frac{{\bf H}_{\rm ext}}{\chi_T^{-1} + \mathcal{D}}.
\end{equation}
Using reasonable parameters, these equations predict a large FC-ZFC splitting in $M(T)/H$, as shown in Fig. \ref{fig4}.
Here it has been assumed that there is a single observation time of about 100 s, which must be a rather crude approximation. Nevertheless, a dramatic FC-ZFC splitting, qualitatively similar to that shown, was observed in experiment by Snyder {\it et al.}~\cite{Snyder}. There appear to be two principal ways in which the experimental result differs from Fig. \ref{fig4}. First, the experimental FC magnetization below the splitting temperature, becomes temperature independent  at a value smaller than the theoretical $M=H/\mathcal{D}$~\cite{dimensional}.  Second,  the experimental splitting temperature ($\sim$0.65 K) for Dy$_2$Ti$_2$O$_7$ is higher than that which can be reasonably justified by Ryzhkin's model. The higher than expected splitting temperature appears to be related to an anomalous slowing down of relaxation seen in ac-magnetization~\cite{Matsuhira-new,Quilliam}, as well as in numerical simulations~\cite{Jaubert}. Possible causes of the experimentally observed slowing down include the constraints imposed by of the Dirac string network~\cite{Jaubert,Quench}, thermal coupling effects~\cite{Slobinsky} 
and a transition in the monopole density~\cite{Ryzhkin-verynew}.
Also, the Wien effect is important in this regime and will play a role in the transient response
~\cite{NatPhys,dimensional}. 

As regards Ryzhkin's prediction~\cite{Ryzhkin} that $\chi_T = 2 \chi_C$, 
a recent theoretical study~\cite{TSF}, using parameters appropriate to ${\rm Ho_2Ti_2O_7}$ spin ice, has shown that there is  
a very slow crossover between $\chi_T = \chi_C$ at high temperature ($\sim 100$ K) to $\chi_T = 2 \chi_C$ in the low 
temperature limit. Experimental measurements appear to be consistent with this prediction~\cite{TSF}. 
This  `Curie law crossover' has not yet been experimentally confirmed for ${\rm Dy_2Ti_2O_7}$. 

\vspace{0.5cm}
\subsection{ac-Magnetization}
For ac-magnetization measurements, Ryzhkin's equation (Eqn. \ref{Ryzhkin}) can be applied, using a demagnetization correction.  As described above, the rate $\nu= 1/\tau_1$, the spin-lattice relaxation rate that arises in the Bloch equations. 

Although Matsuhira {\it et al.} have shown that the relaxation is never a simple exponential at the temperatures of interest~\cite{Mats}, it appears that the characteristic relaxation time does behave according to Ryzhkin's theory. Thus at high temperature we would expect a characteristic relaxation time $\tau = 1/g\nu_0 x$ and this is born out in experiment in the temperature range $> 2$  K for Dy$_2$Ti$_2$O$_7$ where $x \approx 1/2$~\cite{Jaubert}. However at lower temperatures, it is evident that the relaxation rate may not be simply proportional to the monopole density~\cite{Jaubert,Matsuhira-new,Quilliam}.

\vspace{0.5cm}
\section{Neutron Scattering}\label{Neutron}

\subsection{Conventional Neutron Scattering} 

Having accounted for the atomic form factor and assuming sufficiently small energy transfer, the partial differential cross section of conventional neutron scattering ($\sigma''$) is proportional to the imaginary part of the generalised susceptibility:
\begin{equation}
\sigma'' \propto  \frac{kT}{\hbar \omega} {\rm Im}[\chi^{\alpha,\beta}({\bf K}. \omega)].
\end{equation}
Here $\alpha, \beta = x,y,z$ and ${\bf K}$ is the scattering vector. As shown in Ref. \cite{Fennell}, a polarised neutron scattering experiment may be used to isolate the longitudinal ($zz$) susceptibility discussed here by scanning through a Brillouin zone centre perpendicular to the reciprocal lattice vector ${\bf K}_0$. It is particularly useful to use ${\bf K}_0 = (0,0,2)$ in the face centred cubic basis as there is no nuclear Bragg peak at that wavevector~\cite{Fennell}. 

For scans along this direction (which corresponds to a scan across the ``pinch point''~\cite{Fennell}, 
using Eqn. \ref{cqo} and setting ${\bf q = K-K}_0$, we find
\begin{equation}\label{sig}
\sigma''({\rm longitudinal})  \propto   \frac{\kappa T}{(Dq^2+ \nu)^2+ \omega^2}.
\end{equation}
Unfortunately the dynamics of spin ice are generally too slow to test this expression. Instead it is possible to energy integrate and measure in the static approximation whereby the differential quasi-elastic cross section $\sigma'$ is given by:
\begin{equation}
\sigma'({\rm longitudinal}) \propto T\chi(\bf q).
\end{equation}

Using Eqn. \ref{cqo} with $\omega = 0$ we find: 
\begin{equation}\label{cqo2}
\sigma'({\rm longitudinal}) \propto  \frac{\xi^{-2} T \chi_T}{\xi^{-2}+q^2},
\end{equation}
where as already stated, $\xi = a/\sqrt{6gx}$. 

In general $x$ is well approximated by:
\begin{equation}\label{n}
x \approx \frac{\exp(-n(T) J_{\rm eff}/kT)}{1+ \exp(-n(T)J_{\rm eff}/kT)},
\end{equation}
where $J_{\rm eff}$ is the effective exchange parameter for a given spin ice~\cite{BramwellGingras}.  Here the prefactor $n \rightarrow 4$ in the low temperature limit and 
$n \rightarrow 2$ in the high temperature limit as a result of Debye-H\"uckel screening~\cite{Jaubert}.
For Ho$_2$Ti$_2$O$_7$ spin ice $J_{\rm eff}/k \approx 1.8$ K, so there is a regime at intermediate temperature where $\xi \sim \exp(1.8/T)$. In the experiments of Fennell {\it et al.}~\cite{Fennell}, the neutron data along wavevectors perpendicular to 002 were fitted to the sum of a Lorentzian function and a flat background. The inverse Lorentzian width was indeed found to depend on temperature as predicted here ($\exp(1.8/T)$) although its absolute value was much larger than predicted. The flat background was also found to depend on temperature according to Eqn. \ref{n} at high temperatures (with $n=2$ and a correction for `double charge' monopoles).

\subsubsection{Possible Explanation of the Discrepancy}

There are two potential corrections to Eqn. \ref{cqo2} that we did not consider in Ref. \cite{Fennell}. The first stems from 
the modification of  Eqn. \ref{chiqw} to give Eqn. \ref{chinew}, as discussed above. Applying this gives: 
\begin{equation}
\sigma' ({\rm longitudinal}) \propto \frac{\xi^{-2} T \chi_T}{\xi^{-2}+q^2/(1+ Aq^2a^2/3)}.
\end{equation}
The second would account for the wavevector dependent misalignment between ${\bf M} ({\bf q})$ and ${\bf q}$. However, this is a relatively minor correction and is not considered further here. Writing $a = (\sqrt{3}/4) a_{\rm fcc}$ where $a_{\rm fcc}$ is the face centred cubic lattice constant and $q = \sqrt{2} (2\pi/a_{fcc}) h $, we find (for a scan along $hh0$)
\begin{equation}
\sigma' ({\rm longitudinal}) \propto \frac{(\sqrt{8} \pi \xi/a)^{-2} T \chi_T}{(\sqrt{8} \pi \xi/a)^{-2}+h^2/(1+ (A\pi^2/2)h^2)}.
\end{equation}
Using Eqn. \ref{xi} this may also be written:
\begin{equation}
\sigma' ({\rm longitudinal}) \propto \frac{(3gx/4\pi^2)T \chi_T}{(3gx/4\pi^2)+h^2/(1+ (A\pi^2/2)h^2)}.
\end{equation}
These expressions produce a lineshape and temperature dependence that is very similar to that observed in Ref.\cite{Fennell}, in that they incorporate both the apparent Lorentzian (making it appear anomalously sharp) and the flat background, and they also predict the correct temperature dependence in both cases. It would be interesting to compare them in detail to the experimental data.

\subsection{Neutron Spin Echo}

Neutron spin echo measures the intermediate scattering function $S({\bf K},t)$ which is proportional to the frequency Fourier transform of the right hand side of Eqn. \ref{sig}. Thus we predict
\begin{equation}
S({\bf q},t) \sim \kappa T \exp(-\nu_{\bf q} t), 
\end{equation}
with $\nu_{\bf q}$ given by Eqn. \ref{rate0}. A test of this expression would require measuring neutron spin echo for scattering transverse to the pinch point, as above. Experiments so far~\cite{Ehlers1, Ehlers2} have integrated over larger ranges of ${\bf q}$, including transverse fluctuations, and in a temperature range where $\nu_{\bf q} \approx \nu_0$. A temperature independent relaxation rate has been observed~\cite{Ehlers1, Ehlers2}, but for Ho$_2$Ti$_2$O$_7$ this was several order of magnitude faster than that derived by ac-susceptibility on Dy$_2$Ti$_2$O$_7$. Notwithstanding a possible variation between materials it seems likely that the measured relaxation rate is technique dependent, even though its temperature dependence is not. This suggests a high frequency component to the monopole response that is not contained in the present approximations. 

\vspace{0.5cm}
\section{Muon spin relaxation and rotation}

\subsection{Longitudinal Field $\mu$SR}

In a $\mu$SR experiment the muon is self trapped by the lattice distortion it creates. In a dense magnetic oxide like spin 
ice it is therefore prone to distort the local magnetic environment that it is aiming to probe. Despite this, the published results of $\mu$SR experiments are reasonably explained by the monopole model. 

Thus a longitudinal field $\mu$SR experiment on Dy$_2$Ti$_2$O$_7$ was performed by Lago {\it et al.}~\cite{Lago}, who analysed the long time muon depolarisation rate as a measure of the field fluctuation rate. Hence this should have been a measure of $\nu$ or $1/\tau_1$. The temperature dependence of the corresponding relaxation time is indeed very close to that expected, and it seems very likely that the experiment was observing magnetic monopoles. However the magnitude of the relaxation time was an order of magnitude smaller than that inferred from 
ac-magnetization measurements. This would again suggest a high frequency component to the monopole response, as noted above. 

\vspace{0.5cm}
\subsection{Transverse Field $\mu$SR}

If a muon implants into spin ice at a site of large local field, then transverse field $\mu$SR is an uninteresting probe of the spin ice system.  
Hence we will assume that the muon is at a site of zero local field, either within the sample or exterior to the sample, but near the surface. 
While the assumption of zero field sites within the spin ice sample gives a highly consistent description of experiment~\cite{Nature,NatPhys}, 
their existence has been contested on theoretical grounds~\cite{Dunsiger} and the issue has been debated
~\cite{Comment,Claudio,Blundell}.

At sufficiently high temperature (T $>$10 K) we might expect the TF-$\mu$SR dephasing rate $\lambda$ to give a measure of $1/\tau_2$, 
the spin-spin relaxation rate, which may be specified by a BPP type~\cite{BPP} expression: 
$\tau_2^{-1}$: 
\begin{equation}
\frac{1}{\tau_2} = \frac{1}{2\tau_1}+ \frac{\gamma^2  \langle (\delta B^z)^2 \rangle}{\nu'},
\end{equation}
where $\nu'$ is approximately the spin flip rate and $\delta B^z$ is the scale of the fluctuations of the field component parallel to the applied field. As the latter term tends to dominate, we shall only consider this term from now on. 

In the spin ice regime, where $x = \nu/g\nu_0$ is the dominating parameter of the system, the TF-$\mu$SR response is found to have a form that is unfamiliar in the context of $\mu$SR on paramagnets. To explain this it is useful to first consider the dimensional analysis of the problem. 

\subsection{Dimensional Analysis for TF $\mu$SR}

$\mu$SR theory for a simple paramagnet may be formulated in terms of two parameters: $\Delta$ and $\nu'$. Here $\Delta = \gamma \sqrt{\langle B^2\rangle}$ where the right hand term is the instantaneous root mean square field at the muon site, $\gamma$ is the muon gyromagnetic ratio, and $\nu'$ is the relaxation rate of this local field.  In terms of dimensional analysis we would say that $\nu'$ and $\Delta$ constitute two governing parameters, both with the dimensions of $[1/time]$. The quantity of interest in transverse field $\mu$SR is the characteristic rate of muon dephasing, $\lambda$. The formal solution to the problem is: 
\begin{equation}
\lambda = \Delta f\left(\frac{\nu'}{\Delta}\right),
\end{equation}
where $f$ is an undetermined function. 

In the slow fluctuation limit $\nu'/\Delta \rightarrow 0$ and for $\lambda$ to be finite we have $f \rightarrow constant$. 
In the fast fluctuation limit $\Delta/\nu' \rightarrow 0$ and we expect $\lambda\rightarrow 0$. The asymptotic form is in fact linear in the small parameter, $f(1/\epsilon) \sim \epsilon$. The two solutions thus become:
\begin{equation}\label{one}
\lambda_{\rm slow} \sim \Delta,
\end{equation}
\begin{equation}\label{two}
\lambda_{\rm fast} \sim \frac{\Delta^2}{\nu'},
\end{equation}
formulae that are often used for the analysis of $\mu$SR data. 

These formulae may be rationalised by the following heuristic argument. If the field sensed by the muon is approximately static on the muon lifetime,  then the muons precess in phase at a Larmor frequency $\gamma B_a$ where $B_a$ is the applied transverse field, but accumulate a phase difference $\Delta \phi = \Delta t$ in time $t$.  If $1/\lambda$ is equated with the time to dephase by order 1 radian then we obtain $\lambda \sim \Delta$. If, on the other hand, the field jumps randomly at rate $\nu$, with jump magnitude $\Delta$, then the phase difference accumulated between flips is $\Delta \phi = \Delta/\nu'$ 
and the phase undergoes a random walk with end to end distance $\nu t (\Delta/\nu)^2$ in time $t$, yielding Eqn. \ref{two}. 

The case of spin ice is unusual in that there are three, not two, governing parameters. The origin of the third governing parameter is in the thermodynamics of the Coulomb gas in the grand canonical ensemble where the monopole number $N$ is the sole extensive system parameter. We have defined $x=N/N_0$ as a dimensionless monopole density (where $N_0$ is the number of diamond lattice sites) and $\nu_0$ is the temperature independent monopole hop rate. As discussed above, the relaxation rate of the local magnetic field is 
$\nu =  g x \nu_0$ and we may define a scale for the field $\Delta_0$, that depends only on fixed microscopic parameters. 

The formal solution of dimensional analysis can be written:
\begin{equation}
\lambda_{\rm slow} = \Delta_0 f\left(\frac{\Delta_0}{\nu_0}, \frac{\nu}{\nu_0}\right).
\end{equation}
The physical picture we wish to explore is that low temperature ($x \rightarrow 0$) corresponds to slow fluctuations, and high temperature ($x\approx 1$) corresponds to fast fluctuations. Taking the slow fluctuation limit $\nu_0/\Delta_0, \nu/\Delta_0 \rightarrow 0$ now does not necessarily eliminate $\nu/\nu_0 = g x$ from the problem. Whether it does so or not depends on the function $f$. If muons detect monopolar fields only (that is, the longitudinal susceptibility), then we would expect $\lambda$ to go to zero as a power law in $x$, for in the absence of monopoles there should be no dephasing. In contrast, in the fast fluctuation limit $\nu/\nu_0$ does drop out of the problem and we again recover Eqn. \ref{two}. The two solutions appropriate to the detection of monopolar fields are therefore: 
\begin{equation}\label{three}
\lambda_{\rm slow} \sim \Delta_0 \left(\frac{\nu}{\nu_0}\right)^y,
\end{equation}
\begin{equation}\label{four}
\lambda_{\rm fast} \sim \frac{\Delta_0^2}{\nu_0}.
\end{equation}

Thus in the slow fluctuation limit we expect $\lambda \rightarrow \nu^y$, while in the fast fluctuation limit, we expect $\lambda \rightarrow constant$. The former is an unusual result in the context of $\mu$SR and applies to the case where the muons sense only monopolar fields.  

\vspace{0.5cm}
\subsection{TF$-\mu$SR at Low Temperature}

At low temperature the monopole gas is sparse ($ x \ll 1$) and muons that are close to monopoles are rare.
The muon experiment acts to some extent as a spectroscopy, associating different field contributions with different times of observation. Hence to use the average field may not be quite correct. The muon signal at long times measures only typical muons, which, are far from magnetic monopoles. The typical distance of a muon to a monopole is approximately
$r^{\ast}/a \approx x^{-1/3}$ and the field sensed by the muon is $
|B| \approx B_0 x^{2/3}$. Since this field is random in direction we get the same result for the mean square field as above, but with the 
exponent $4/3$ on $x$ instead of $1$. In general we might expect the apparent mean square field to be given by the equation $\langle B \rangle^2 =  B_0^2x^y$, with $y\approx 1$.

In this limit the Debye length $l_D$ is very large and the conductivity $\kappa$ is very small. Although $\kappa$ scales with $x$, if $y < 2$ 
then $\gamma \sqrt{\langle B^2\rangle }$ is always larger than it, and the fields are quasistatic (here $\gamma$ is the muon gyromagnetic ratio). If we approximate the fields as completely static on the muon lifetime, then the muons sense a $z-$component of the local field that is of the order of the root mean square field. The field sensed by the muons is approximately $B_0 \pm \sqrt{\langle B^2\rangle }$ and the muons precess coherently at a Larmor frequency $\gamma B_0$ but are dephased by the spread in local fields. Introducing $\Delta = \gamma  \sqrt{\langle B^2\rangle }$ the spread of phases accumulated in time $t$ is 
\begin{equation}
\Delta \phi  = t \Delta. 
\end{equation}
The dephasing time $1/\lambda$ is equated with the time taken for $\Delta \phi$ to become of order one radian, with the result \footnote{Henley~\cite{Henley2} has considered relaxation functions $G(t)$ for nmr for nuclei in sites of zero local field, and in the dilute monopole limit has shown that $G(t) = \exp \left(-n (t/\tau_i)^{\beta}\right)$ where $\beta$ is a positive exponent of order unity and $\tau_i (i=1,2)$ is a characteristic timescale for longitudinal (1) and transverse (2) relaxation. The results obtained here (for zero local field sites) are consistent with the $\mu$SR relaxation function taking this general form.} :
\begin{equation} 
\lambda^{\rm low~T} = \Delta \approx \gamma B_0 x^{y/2}.
\end{equation} 
Hence $\lambda$ is 
\begin{equation}
\lambda^{\rm low~T} = \gamma B_0 \left(\frac{\nu}{g\nu_0}\right)^{y/2} = \gamma B_0 \left(\frac{\kappa}{g\chi_T \nu_0}\right)^{y/2}.   
\end{equation}

The muon dephasing function depends on the actual field distribution. However it is always of the form: 
\begin{equation}
P = P(\lambda t).
\end{equation}
This form (with $\lambda \propto \nu$) was assumed in Ref. \cite{Nature} and gave a highly consistent description of experiment. Although this applies the current ideas in the Wien effect regime, one would expect this to be reasonable on the grounds of the dimensional arguments given above. Note also that the method of Ref. \cite{Nature} is insensitive to the precise form of the local field distribution. The typical value of $\lambda$ observed in Ref. \cite{Nature} was of the order $10^5$ s$^{-1}$. For DTO spin ice $\gamma B_0 \approx 10^8$ s$^{-1}$, so a $\lambda$ of $10^5$ s$^{-1}$ corresponds to x = $10^{-6}$ if $y = 1$ and the monopole field at a typical muon site is about $10^{-3}$ T. The temperature at which the monopole density is expected to fall to this value is 0.3 K, which is consistent with the observations of Ref. 
\cite{Nature}.  
 
\vspace{0.5cm}
\subsection{TF$-\mu$SR at High Temperature}
 
In the high temperature limit $x$ becomes of order unity so $\nu \approx \nu_0$. Thus as we pass from low to high temperature, monopoles hopping at a rate $\nu_0$ located near to the muon become increasingly important, but as remarked above, these monopoles cannot be distinguished from spins, and we return to a model of spin flipping at rate $\nu_0$. In this case the ordinary equations of $\mu$SR apply. 

\section{Conclusion}

The main conclusion of the present work is that magnetic monopoles in spin ice largely determine the longitudinal response of the system. The sole system variable for both static and dynamic response is the dimensionless monopole density $x$, which is determined in a complex way by the four fixed parameters of the problem: $a,Q,\mu$ and $\nu_0$. In contrast, the transverse response does not directly mirror monopole correlations.

The main theoretical results of this paper are contained in Eqns. \ref{x}, \ref{newcurrent}, \ref{Landau}, \ref{nuq}, \ref{xi}, \ref{Gfield}, \ref{rate1}, \ref{far}, 
\ref{near}. 

Temporal and spatial correlations are linked by $x$ and the Eqns. \ref{x} and \ref{xi} combine to establish a dynamic scaling relation: 
\begin{equation}
\nu \sim \xi^{-z}
\end{equation}
with $z = 2$, as would be expected for a problem of Brownian motion. It follows (see  Eqn. \ref{nuq}) that there is a dispersion of relaxation rates on all scales from the monopole hop rate $\nu_0$ to the magnetization relaxation rate $\nu$. Some evidence has been noted to suggest that field fluctuations relax more quickly than spin fluctuations (see Eqns. \ref{nuq}, \ref{rate1}) but more work is needed to establish this. 

The exponents $\nu$ and $z$ defined in this way, and the correlation length $\xi$, are not conventional quantities as they reflect monopole rather than spin correlations. The spin correlations obey static scaling in the following sense. The correlation function $g(r)$, being pseudo~dipolar~\cite{YA,Henley} decays as $g(r)\sim r^{-3}$. Applying the scaling relation $g(r)~r^{-(d-2+\eta)}$ we find $\eta = 2$. As the susceptibility diverges as $1/T$, the susceptibility exponent $\gamma$ takes the value $\gamma=1$. Applying the scaling relation $\nu = \gamma/(2-\eta)$ we find $\nu$ is infinite, meaning that the spin-spin correlation length remains finite at all temperatures. Thus $T=0$ marks an unusual critical point with algebraic decay of spin correlations, a divergent spin susceptibility, but a non-divergent spin correlation length. It is interesting to observe however, that the monopole correlation length does diverge at $T=0$. 

Free energy functionals for the magnetization and field fluctuations have been derived (Eqns. \ref{Landau}, \ref{Gfield}) and shown to relate closely to Eqn. \ref{newcurrent}, previously stated by Ryzhkin and Ryzhkin~\cite{Ryzhkin-new}. In future work it would be interesting to express these as functionals of the density ($x$) and to add further terms to account for energy fluxes in the system, as well as Wien dissociation, which both play a role at low temperature~\cite{Nature, NatPhys,Klemke}. 

The generalised susceptibility (Eqn. \ref{nuq}) at the level of Ryzhkin's  description~\cite{Ryzhkin} has been derived, as well as the field fluctuation at the level of Debye-H\"uckel theory~\cite{CMS2}. The latter was used to calculate the longitudinal field fluctuation at a point in the system (Eqns. \ref{far}, \ref{near}), which may be compared to established results for electrolytes~\cite{Oosawa}.  The expressions for the mean square field distribution have been used to show that a point probe such as a muon at a `spin free' site (either inside the sample or just outside) will give a direct measure of the monopole density as assumed in Ref. ~\cite{Nature}.

It has been shown that according to the electrolyte theory, a non-equilibrium population of monopoles is always frozen into the sample, regardless of the rate of cooling. However in sufficiently weak magnetic field there is never a phonon bottleneck. The former effect should generally be considered when treating low temperature experimental data. 

In general the theory discussed here works qualitatively well for real spin ice materials, capturing the temperature, wavevector and time dependence of a diverse range of experimental responses. However, there are three clear discrepancies. First, while the temperature and wavevector dependence of the neutron scattering cross section are well accounted for, the amplitude of the correlation length is not, being an order of magnitude longer in experiment compared to theory~\cite{Fennell}; a possible explanation of this has been proposed here. Second, while different experiments~\cite{Snyder, Mats, Ehlers1,Ehlers2, Lago} agree on the temperature dependence of the relaxation rate, they exhibit a wide range of relaxation rates: it appears that there is a high frequency response, not accounted for in the hopping model. Third, the ac-susceptibility relaxation in the high temperature limit is more strongly dispersed~\cite{Mats} than predicted by the simple approximations discussed here. 

It seems very unlikely that the monopole theory will have to be abandoned to explain these discrepancies. More likely it needs to be refined. In addition to the possible revision of the neutron scattering line shape discussed above, one might also need to consider the effect of quantum fluctuations~\cite{Shannon,Benton} or minor terms in the spin ice Hamiltonian~\cite{Yavorskii}. Also, microscopic factors affecting the rate of local spin flipping or monopole hopping probably remain to be identified. However, it should also be emphasised that the most distinctive aspect of Coulombic correlation - the tendency to Bjerrum pairing - has not been accounted for here, or in other `high temperature' theories, but will certainly play a role. Thus Bjerrum pairs have been argued to be important in the low temperature non equilibrium regime~\cite{NatPhys} and have been identified in specific heat measurements~\cite{Zhou}. Finally, the Wien effect~\cite{Onsager}, though weak at the `high' temperatures considered here, still exists in a screened form~\cite{Liu},  and should be accounted for in a more accurate description. Although there is much work to be done, it is clear that the monopole theory of spin ice~\cite{CMS,Ryzhkin} is a remarkably simple and effective description of a complex condensed matter system. 

\ack{It is a pleasure to thank P. Holdsworth, S. Banks, S. Giblin, J. Bloxsom, L. Bovo, G. Aeppli, V. Kaiser, C. Castelnovo, R. Moessner, S. Blundell and I. Ryzhkin for very useful discussions and comments.}

\end{document}